\pdfoutput=1
\documentclass[aps,pre,10pt,twocolumn,byrevtex,footinbib,bibnotes,showpacs,showkeys,longbibliography]{revtex4-2}
\usepackage{microtype}
\usepackage{graphicx}
\usepackage{amsmath}
\usepackage{amssymb}
\usepackage{color}
\definecolor{myblue}{rgb}{0.153,0.322,0.706}
\usepackage[colorlinks,linkcolor=myblue,urlcolor=myblue,citecolor=myblue,breaklinks=true]{hyperref}
\usepackage{algorithm}
\usepackage[noend]{algpseudocode}
\algrenewcommand{\algorithmiccomment}[1]{$\vartriangleright$ #1}
\algrenewcommand{\algorithmicreturn}{\textbf{Return: }}
\algnewcommand\algorithmicinput{\textbf{Input: }}
\algnewcommand\Input{\State \algorithmicinput}

\newcommand{\be}{\begin{equation}}
\newcommand{\ee}{\end{equation}}

\newcommand{\ra}{\rightarrow}

\newcommand{\p}{\partial}

\newcommand{\tPi}{\tilde\Pi}
\newcommand{\hX}{\hat X}

\newcommand{\ER}{Erd\"os--R\'enyi~}

\begin{document}

\title{Adaptive power method for estimating large deviations in Markov chains}

\author{Francesco Coghi}
\email{francesco.coghi@su.se}
\affiliation{Nordita, KTH Royal Institute of Technology and Stockholm University, Stockholm, Sweden}

\author{Hugo Touchette}
\email{htouchette@sun.ac.za}
\affiliation{Department of Mathematical Sciences, Stellenbosch University, Stellenbosch, South Africa}

\date{\today}

\begin{abstract}
We study the performance of a stochastic algorithm based on the power method that adaptively learns the large deviation functions characterizing the fluctuations of additive functionals of Markov processes, used in physics to model nonequilibrium systems. This algorithm was introduced in the context of risk-sensitive control of Markov chains and was recently adapted to diffusions evolving continuously in time. Here we provide an in-depth study of the convergence of this algorithm close to dynamical phase transitions, exploring the speed of convergence as a function of the learning rate and the effect of including transfer learning. We use as a test example the mean degree of a random walk on an \ER random graph, which shows a transition between high-degree trajectories of the random walk evolving in the bulk of the graph and low-degree trajectories evolving in dangling edges of the graph. The results show that the adaptive power method is efficient close to dynamical phase transitions, while having many advantages in terms of performance and complexity compared to other algorithms used to compute large deviation functions.
\end{abstract}

\maketitle

\section{Introduction}

The application of ideas and techniques from large deviation theory in nonequilibrium statistical physics has led to many new insights about the transport properties of steady-state nonequilibrium systems \cite{derrida2007,bertini2007,bertini2015b} and other interesting phenomena such as dynamical phase transitions \cite{garrahan2007,hedges2009,garrahan2009,garrahan2010,espigares2013,bunin2013,tsobgni2016b,lazarescu2017}, fluctuation symmetries \cite{gallavotti1995,kurchan1998,lebowitz1999,harris2007}, and dissipation bounds \cite{barato2015b,pietzonka2015,gingrich2016,gingrich2017,li2019}. Within this theory, the fluctuations of observable quantities, such as particle currents and kinetic activities, are characterised by a function or potential, called the rate function, giving the rate of decay of the probability distribution of an observable in the limit where the observation time or volume of the system considered goes to infinity, similarly to the thermodynamic limit of equilibrium statistical mechanics \cite{touchette2009}.

Finding the rate function of a given observable and Markov process modelling a nonequilibrium system is difficult in general, as it involves solving a spectral problem or a related optimization problem whose dimension increases with the size of the process considered \cite{touchette2017}. As a result, many numerical techniques have been devised over the years to compute the rate function by either solving the spectral or optimization problems using discretization or projection methods \cite{ray2018,banuls2019,casert2021,causer2021,das2021b} or by simulating the underlying process in a way that goes beyond direct simulation \cite{juneja2006,asmussen2007,bucklew2004,guyader2020}, which is obviously inefficient for sampling large fluctuations and rare events in general. On the simulation side, techniques such as cloning \cite{giardina2006,lecomte2007a,nemoto2016,perez2019,angeli2019}, splitting \cite{cerou2007,dean2009,cerou2019b,brehier2019}, and importance sampling \cite{kundu2011,ray2018b,klymko2018,ray2020} have also been developed and used with good results, although more work is needed to efficiently deal with complex systems involving many interacting particles. Recently, machine learning tools have been used to address this issue \cite{oakes2020,rose2021,das2021,yan2022}.

In this paper, we focus on a simulation method that adaptively estimates or learns the rate function using a stochastic variation of the power method for computing eigenvalues that combines techniques from importance sampling and stochastic approximations. This method or algorithm, referred to as APM for \emph{adaptive power method}, was introduced in a series of works on stochastic control theory \cite{borkar2004,ahamed2006,basu2008} and was applied in that context to estimate the likelihood of rare events arising, for example, in communication networks and portfolio optimization. Recently, APM was also applied to continuous-state and continuous-time processes, which require a function representation or projection of their state \cite{ferre2018}.

Here, we investigate the performance of APM when applied to systems showing dynamical phase transitions (DPTs), i.e., phase transitions in the fluctuations, signalled by singularities in the rate function or its dual, the scaled cumulant generating function \cite{garrahan2007,hedges2009,garrahan2009,garrahan2010,espigares2013,bunin2013,tsobgni2016b,lazarescu2017}. The observable and process that we consider as a test case is the mean degree of a uniform random walk on an \ER (ER) random graphs, which appears to show a DPT separating high-degree trajectories of the random walk evolving in the bulk of the random graph and low-degree trajectories evolving in dangling edges of the graph \cite{bacco2015b,bacco2015,coghi2018b}. For this system, we study the convergence and accuracy of APM as a function of the system size and learning rate, and also look at the effect of using transfer learning when initializing the algorithm.

Studying DPTs numerically is generally difficult because of the need of simulating large systems and the presence of slowing down effects \cite{lecomte2007a,nemoto2016,perez2019}. Our results show that the performance of APM is not affected by DPTs, as it includes importance sampling, which guides the estimation in the relevant region of the process where the phase transition takes place. This importance sampling step can be included in other algorithms, such as cloning, but involves a more complicated feedback mechanism requiring a large number of parallel simulations \cite{nemoto2016}. By comparison, APM estimates the rate function in a direct way using only one simulated trajectory of a process that is gradually controlled towards the rare event or fluctuation of interest. This makes APM a simple and efficient algorithm for studying DPTs and long-time large deviations in general. Other advantages of APM are discussed together with some technical improvements for dealing with large systems.

\section{Large deviations of Markov chains}

We review in this section the spectral method used in large deviation theory to obtain the rate function of time-integrated functions of Markov processes, which is the starting point of APM. For simplicity, we consider the case of homogenous Markov chains evolving in discrete time, which are used to model various processes, including population dynamics, queues, chemical reactions, molecular motors and other physical systems having a discrete number of states \cite{gardiner1985,grimmett2001,jacobs2010}.

\subsection{Markov chain model}

We denote by $(X_\ell)_{\ell=1}^n$ the sequence of states or \emph{trajectory} of a Markov chain evolving in a state space $\Gamma$ over $n$ time steps. We assume that $\Gamma$ is discrete and finite and that the transition matrix $\Pi$, whose element $\Pi(i,j)$ gives the probability of going from the state $i$ to $j$ over one time step, is irreducible and aperiodic, so the Markov chain is ergodic \footnote{We use the notation $\Pi(i,j)$ instead of the more usual $\Pi_{ij}$ to denote the $(i,j)$ entry of $\Pi$ becausse other subscripts will be used after.}. As a result, it has a unique stationary probability distribution $p^*(i)$, $i\in \Gamma$, satisfying $p^*\Pi = p^*$ in (row) vector notation, which gives the fraction of time spent in each state in the long-time limit.

For this Markov model, we are interested to study the fluctuations of time-additive quantities, costs or observables having the form
\begin{equation}
\label{eq:Cost}
C_n = \frac{1}{n} \sum_{\ell=1}^n f(X_\ell),
\end{equation}
where $f$ is some real function of the state. This random variable can represent, for example, the mean energy of a system transitioning randomly between energy levels or the fraction of time spent in some state. To be more general, one can also consider observables defined with a function $g(X_\ell,X_{\ell+1})$ instead of $f(X_\ell)$ so as to include jump contributions to $C_n$. This is commonly done in physics, for example, when considering particle and probability currents, as well as thermodynamic quantities such as the entropy production \cite{seifert2018}. The results in this case follow with minor modifications of the results obtained here with $f(X_\ell)$ \cite{chetrite2014}.

\subsection{Large deviations}

The theory of large deviations predicts that the probability distribution of $C_n$, denoted by $P_n(c)$, scales with $n$ according to 
\begin{equation}
\label{eq:Rate}
P_n(c) \sim e^{-n I(c)},
\end{equation}
with sub-exponential corrections in $n$ \cite{shwartz1995,dembo1998,hollander2000}. Thus the problem that we consider is that of estimating the decay function $I(c)$ so as to get information about the fluctuations of $C_n$ to leading order in $n$. This function is called the \emph{rate function} and is defined by the limit
\begin{equation}
I(c) = \lim_{n\ra\infty} -\frac{1}{n}\ln P_n(c).
\end{equation}
We refer to den Hollander \cite{hollander2000} for a more mathematical and rigorous definition of the rate function based on probability measures and the large deviation principle.

The rate function is positive and is equal to $0$ for ergodic Markov chains only for the typical value of $C_n$ corresponding to the ergodic expectation
\begin{equation}
\label{eq:TypCost}
c^* = \sum_{i\in\Gamma} p_i^* f(i).
\end{equation}
Indeed, by the ergodic theorem, $C_n$ converges in probability to $c^*$ as $n\ra\infty$, so that $P_n(c)$ concentrates around $c^*$ in that limit. The scaling \eqref{eq:Rate} extends this result by showing that $P_n(c)$ concentrates \emph{exponentially} around $c^*$, meaning that fluctuations away from the typical value $c^*$ are exponentially unlikely with $n$.

The rate function can be obtained using different methods from large deviation theory \cite{dembo1998}.  The most common proceeds by considering the scaled cumulant generating function (SCGF), defined by the limit
\begin{equation}
\label{eq:SCGFDef}
\Psi(s) = \lim_{n \ra \infty} \frac{1}{n} \ln E [ e^{s n C_n}],
\end{equation}
where $s \in \mathbb{R}$ and $E[\cdot]$ denotes the expectation. For ergodic Markov chains, the SCGF is known \cite{dembo1998} to be given by the logarithm  of the dominant eigenvalue of a transformation of the transition matrix $\Pi$, referred to as the \emph{tilted matrix}, defined by
\begin{equation}
\label{eq:TiltedMatrix}
\tPi_s (i,j)= \Pi(i,j) e^{s f(i)}.
\end{equation}
As a result, we have 
\begin{equation}
\label{eq:SCGFMarkov}
\Psi(s) = \ln \zeta_s,
\end{equation}
where $\zeta_s$ denotes the unique (Perron--Frobenius) dominant eigenvalue of $\tPi_s$, which is a positive matrix. From this point, one then uses the G\"artner--Ellis theorem \cite{dembo1998} to obtain $I(c)$ as the Legendre transform of $\Psi(s)$, given by
\begin{equation}
\label{eq:Legendre}
I(c) = s_c c - \Psi(s_c),
\end{equation}
where $s_c$ the unique solution of
\begin{equation}
\label{eq:LegDual}
c = \Psi'(s).
\end{equation}
This has an obvious similarity with the Legendre transform of thermodynamics connecting the entropy and free energy. For this reason it is common to interpret the parameter $s$ of the SCGF as a Legendre parameter or ``temperature'' conjugated or dual to $C_n$ \cite{touchette2009}.

\subsection{Effective Markov chain}

The result above shows that the problem of finding the rate function reduces to solving a spectral problem, namely,
\begin{equation}
\label{eq:EigenProblem}
\tPi_s r_s = \zeta_s r_s ,
\end{equation}
for the dominant eigenvalue $\zeta_s$ and its corresponding eigenvector $r_s$ with components $r_s(i)$, $i\in\Gamma$. The knowledge of this eigenvector is also important in large deviation theory, as it provides a way to understand how a given fluctuation $C_n=c$ away from $c^*$ is created by a \emph{modified Markov chain} with transition matrix
\begin{equation}
\label{eq:DrivenTransition}
\Pi_s (i,j)= \frac{\tPi_s(i,j) r_s(j)}{\zeta_s r_s(i)}=\frac{e^{sf(i)}\Pi(i,j) r_s(j)}{\zeta_s r_s(i)}.
\end{equation}
Since $r_s>0$ and $\zeta_s> 0$, $\Pi_s$ is a well-defined non-negative matrix, which is also stochastic and ergodic \cite{miller1961}. As a result, it defines another ergodic Markov chain on $\Gamma$, which is known to be equivalent to the original Markov chain conditioned on the event $\{C_n=c\}$ in the long-time limit where $n\ra\infty$, provided that $s$ is chosen according to the duality relation \eqref{eq:LegDual} \cite{chetrite2013,chetrite2014,chetrite2015}. Hence, the modified Markov chain can be interpreted as an effective Markov model that describes the subset of paths of the original Markov chain that lead to the fluctuation $C_n = c$.

From a simulation point of view, the Markov chain described by $\Pi_s$ can also be seen an exponential tilting of the original Markov chain, which can be used to efficiently sample the event $\{C_n=c\}$ using importance sampling \cite{asmussen2007,bucklew2004,guyader2020}. This follows because the typical value of $C_n$ in the modified Markov chain is $c_s=\Psi'(s)$, so choosing $s$ according to \eqref{eq:LegDual} gives us $c_s=c$ \cite{chetrite2015}. In other words, what is a rare event for the original Markov chain is transformed into a typical event for the modified Markov chain.

This can be applied to estimate the SCGF. The exponential expectation $E[e^{snC_n}]$ entering in the definition of the SCGF is dominated by rare fluctuations of $C_n$ centered around $c_s$ \cite{touchette2009}. Sampling those fluctuations with the original Markov chain is inefficient when $s\neq 0$, whereas sampling them with the modified process is efficient \cite{chetrite2015}. This is important for understanding APM. 

\section{Adaptive power method}

The dominant eigenvalue $\zeta_s$ and its corresponding eigenvector $r_s$ can be obtained numerically by diagonalizing $\tPi_s$, but this method is not efficient for large systems, as it generally requires $O(|\Gamma|^3)$ steps for a system of size $|\Gamma|$. Since we need only the dominant eigenvalue, we can use instead the power method, which proceeds by choosing an initial (positive) vector $r^{(1)}$ and by calculating successive approximations of $r_s$ in a recursive way using the (column) matrix product
\begin{equation}
\label{eq:RightEigvApproxPow}
r^{(m+1)} = \tPi_s r^{(m)}.
\end{equation}
From this update, the dominant eigenvalue is then approximated as
\begin{equation}
\label{eq:EigvApproxPow}
\zeta^{(m+1)} = \frac{r^{(m+1)} (i)}{r^{(m)}(i)},
\end{equation}
where $i$ is any component in $\Gamma$. 

Provided that the norm of $r^{(m)}$ is kept constant at every iteration by including a normalization step, as explained in detail below, we have $r^{(m)}\ra r_s$ up to an arbitrary multiplicative constant and $\zeta^{(m)}\ra \zeta_s$ in the limit where $m\ra\infty$. In particular, we can divide $r^{(m)}$ by its maximum component, located say at $i_0$, before the next matrix product to obtain
\be
\zeta^{(m+1)} = r^{(m+1)}(i_0).
\ee
This corresponds to choosing the infinity norm for the normalization.

The complexity of the power method is $O(|\Gamma|^2)$, since it is based on repeated matrix products, which is better than diagonalization but still inefficient for large systems. The idea of APM is to apply the power method in a statistical way by observing that the action of $\tPi_s$ is a conditional expectation over one step of the Markov chain:
\begin{eqnarray}
(\tPi_s r)(i) &=&\sum_{j\in\Gamma} e^{s f(i)} \Pi(i,j) r(j)\nonumber \\
&=& e^{s f(i)}E[r(X_{\ell+1})|X_\ell = i].
\label{eq:exppois1}
\end{eqnarray}
This suggests estimating the expectation using Monte Carlo simulations by drawing samples of the Markov chain. In this way, we avoid calculating the matrix product over all the states, focusing instead on those that are actually visited.

\begin{figure}
\begin{algorithm}[H]
\caption{Adaptive power method (APM)}
\label{alg:APM} 
\begin{algorithmic}
	\State{\textbf{Data:} Transition matrix $\Pi$, initial state $x_1$, Lagrange parameter $s$, number of time steps $n$, learning sequence $a_\ell$.}
    	\State{\textbf{Initialisation:} $r^{(1)}=1$, $\zeta^{(1)}=1$, $\Pi_s^{(1)} = \Pi$, $\gamma^{(1)}=1$}
	\For{$\ell=1,\dots, n-1$}
		\State{Draw $x_{\ell+1}$ from $x_{\ell}$ with probability $\Pi_s^{(\ell)}(x_{\ell},\cdot)$} 
		
		\State{Update $r^{(\ell)}$ to $r^{(\ell+1)}$ at  $i=x_{\ell}$ using \eqref{eq:update1}}
		\State{Locate max of eigenvector: $i_0 \leftarrow\arg\max_k r^{(\ell+1)}(k)$}
		\State{Update eigenvalue: $\zeta^{(\ell+1)}\leftarrow r^{(\ell+1)}(i_0)$}
		\State{Compute normalizing factor $\gamma^{(\ell+1)}$ at $i=x_{\ell}$ using \eqref{eq:modnorm1}}
		\State{Update $\Pi_s^{(\ell+1)}$ at $i=x_{\ell}$ and $j=x_{\ell+1}$ using \eqref{eq:modmat2}}
	\EndFor
\State{{\bfseries return:} $\hat\Psi_n(s) = \ln \zeta^{(n)}$}.
\end{algorithmic}
\end{algorithm}
\end{figure}

This approach for calculating the SCGF was proposed by the group of Borkar \cite{borkar2004,ahamed2006,basu2008}, who suggested two different simulation schemes for estimating the expectation \eqref{eq:exppois1}: a concurrent or \emph{synchronous} scheme in which many copies of the Markov chain are simulated starting from the state $i$, and an \emph{asynchronous} scheme in which the expectation is estimated as an ergodic average using a single trajectory of the Markov chain. The latter scheme appears to us simpler and is therefore the one that we consider here. It is summarised in Algorithm~\ref{alg:APM} and involves three steps:

1- \emph{Importance sampling}: The estimation of the expectation in \eqref{eq:exppois1} using samples of the Markov chain $(X_\ell)_{\ell=1}^n$ becomes inefficient as more and more steps of the power method are taken, since this method tries to estimate the exponential expectation in the definition of the SCGF \cite{ferre2018}, which cannot be estimated efficiently, as mentioned, from trajectories of the original Markov chain. Instead, we must generate samples from the modified Markov chain, so as to write the same expectation with importance sampling as
\begin{eqnarray}
\label{eq:drivenexp1}
(\tPi_s r)(i) &=&\sum_{j\in\Gamma} e^{s f(i)} \Pi_s(i,j) \frac{\Pi(i,j)}{\Pi_s(i,j)} r(j)\nonumber \\
&=& e^{sf(i)} E[r(\hX_{\ell+1}) R_s(\hX_\ell,\hX_{\ell+1})|\hX_\ell = i],
\end{eqnarray}
where $\hX_\ell$ denotes the modified Markov chain with transition matrix $\Pi_s$ given by \eqref{eq:DrivenTransition} and
\be
R_s(i,j) = \frac{\Pi(i,j)}{\Pi_s(i,j)}
\label{eqlike1}
\ee
is the likelihood factor that corrects for the fact that the expectation involves $\hX_\ell$ rather than $X_\ell$. 

In the algorithm, states of the modified Markov chain are generated from the current estimates $\zeta^{(\ell)}$ and $r^{(\ell)}$ of the dominant spectral elements at time $\ell$ \footnote{Spectral elements are updated at the same time as states of the modified Markov chain are generated, so that $m=\ell$.}, resulting in the following estimate of the modified transition matrix:
\be
\label{eq:modmat1}
\Pi_s^{(\ell)} (i,j)= \frac{e^{sf(i)}\Pi(i,j)r^{(\ell)}(j)}{\zeta^{(\ell)} r^{(\ell)}(i) \eta^{(\ell)}(i)},
\ee
which includes an additional normalization factor $\eta^{(\ell)}(i)$, compared to \eqref{eq:DrivenTransition}, because $r^{(\ell)}$ is not exactly the dominant eigenvector of $\tPi_s$. As a result, we must introduce this factor so that
\be
\sum_{j\in\Gamma} \Pi_s^{(\ell)}(i,j) =1
\ee
for all  $i\in\Gamma$. It is easy to check that imposing this normalization is equivalent to rewriting $\Pi_s^{(\ell)}$ as
\be
\label{eq:modmat2}
\Pi_s^{(\ell)}(i,j) = \frac{\Pi(i,j) r^{(\ell)}(j)}{\gamma^{(\ell)}(i)},
\ee
where
\be
\label{eq:modnorm1}
\gamma^{(\ell)}(i) = \sum_{j\in\Gamma} \Pi(i,j) r^{(\ell)}(j)
\ee
is a different normalization factor. 

2- \emph{Stochastic approximation}: The expectation in \eqref{eq:drivenexp1} is not calculated as a matrix product, as mentioned before, but evaluated pointwise on realisations of the modified Markov chain. Considering, the power method update in \eqref{eq:RightEigvApproxPow} with the result \eqref{eq:modmat2} means that the eigenvector is updated stochastically as
\be
\label{eq:recur1}
r^{(\ell+1)}(i) = \frac{e^{sf(i)}\gamma^{(\ell)}(i)}{\zeta^{(\ell)}}
\ee
at the realisation $i$ of $X_\ell$ at time $\ell$.

3- \emph{Annealing}: Applying \eqref{eq:recur1} recursively on random samples of the modified Markov chain will not result in a convergent estimation of $r_s$ or $\zeta_s$, since $\gamma^{(\ell)}$ always changes randomly as new states are visited. To filter this ``noise'', we can use an annealing scheme whereby the update of $r_s$ is progressively reinforced according to
\be
\label{eq:update1}
r^{(\ell+1)}(i)=(1-a_\ell) r^{(\ell)}(i) + a_\ell \frac{e^{s f(i)}\gamma^{(\ell)}(i)}{\zeta^{(\ell)}}
\ee
using a sequence $(a_\ell)_{\ell\geq 1}$, called the \emph{learning sequence}, that vanishes as $\ell\ra\infty$ in such a way that
\be
\sum_{\ell=1}^\infty a_\ell=\infty,\qquad \sum_{\ell=1}^\infty a_\ell^2 <\infty.
\ee 
These two conditions are known to be sufficient for the stochastic approximation to converge \cite{borkar2008}, and are meant intuitively to balance the need between exploration and exploitation. In practice, other decreasing sequences can be used that satisfy 
\be
a_\ell=\ell^{-\alpha}
\ee 
with $\alpha>0$. We call the exponent $\alpha$ the \emph{learning rate}.

In the end, the steps combine to work as detailed in Algorithm~\ref{alg:APM}: From the state $x_{\ell}$ reached at the $\ell$-th iteration, we generate a new random state $x_{\ell+1}$ with probability $\Pi_s^{(\ell)}(x_\ell,\cdot)$, where $\Pi_s^{(\ell)}$ is the current estimate of the modified transition matrix given by \eqref{eq:modmat1} or \eqref{eq:modmat2}. From these two states, we then update the eigenvector at the location $i=x_\ell$ according to \eqref{eq:update1} and the modified Markov matrix at the locations $i=x_{\ell}$ and $j=x_{\ell+1}$ according to \eqref{eq:modmat2}, having computed the normalizing factor in \eqref{eq:modnorm1}. This can be done in $O(|\Gamma|)$ steps, since only one entry of $\Pi_s^{(\ell)}$ is modified, so only one of its rows needs to be normalized. The complexity of one step of the APM is thus reduced compared to the power method from $O(|\Gamma|^2)$ to $O(|\Gamma|)$.

The result of the algorithm is an eigenvalue estimate $\hat\Psi_n(s)$ of the SCGF $\Psi(s)$ for one value of $s$, which can be repeated so as to obtain an interpolation of this function over a range of values. Alternatively, the SCGF can be estimated using the time-additive estimator
\be
\bar\Psi_n(s) = s C_{n,s} - K_{n,s},
\label{eq:psicostave1}
\ee
where $C_{n,s}$ is the observable value obtained by running the algorithm for the parameter $s$, which converges in probability to $c_s=\Psi'(s)$ as $n\ra\infty$, and 
\be
K_{n,s} =-\frac{1}{n} \sum_{\ell=1}^{n-1} \ln R_s(x_\ell, x_{\ell+1})
\label{eq:costest1}
\ee
is an additive cost associated with the likelihood ratio, which is also estimated using the trajectory of the modified Markov chain generated by the algorithm. The form of $\bar\Psi_{n}(s)$ follows from a control interpretation of large deviation functions for Markov processes \cite{chetrite2015} and is advantageous for computing error bars.

To obtain the rate function, we can in principle take the Legendre transform of either estimator of the SCGF. However, although $\Psi(s)$ is convex by definition \cite{touchette2009}, neither $\hat\Psi_n(s)$ nor $\bar\Psi_n(s)$ is found in simulations to be convex because of statistical errors (see Fig.~\ref{figldfct1}), so taking their Legendre transform would yield spurious results. To avoid this problem, which is not specific to our model \cite{ferre2018}, it is best to estimate $I(a)$ using the additive cost $K_{n,s}$, which converges in probability to $I(c_s)$ as $n\ra\infty$ \cite{chetrite2015}. By computing the pair $(C_{n,s}, K_{n,s})$ for varying $s$ values, we then obtain an approximation of the rate function in parametric form as $(c_s, I(c_s))$ \cite{ferre2018}. An advantage of this method is that, as for the SCGF, error bars can be computed directly.

\section{Application}

We investigate in this section the performance of the APM algorithm for a system showing a DPT, specifically, a simple random walk evolving on an ER random graph, for which the observable is taken to be the mean of the degrees of the nodes visited by the random walk \cite{bacco2015b,bacco2015,coghi2018b}. We define this model next and review recent results about its large deviations and the existence of a first-order DPT in the thermodynamic limit. We then reproduce these results using APM, testing different annealing schedule and the effect of transfer learning.

\subsection{Model}

We consider a random walk evolving on an undirected graph $G$ with $N$ vertices and $M$ edges or links, determined by the adjacency matrix
\be
A_{ij} = \left\{
\begin{array}{lll}
1 & & i,j \textrm{ are connected}\\
0 & & \textrm{otherwise.}
\end{array}
\right.
\ee
The graph is chosen randomly from the ER random ensemble of graphs in which edges are chosen in a binomial way over the $N(N-1)/2$ possible edges with probability $p=\kappa/N$, where $\kappa>1$ \footnote{This scaling of the link probability $p$ defines the sparse regime of the ER ensemble, in which most graphs have a connected component of size proportional to $N$ and have a degree distribution that becomes Poissonian with expectation $\kappa$ in the ``thermodynamic limit'' $N\ra\infty$.}. The random walk itself is determined by the transition probabilities
\be
\Pi(i,j) = \frac{A_{ij}}{k_i}
\ee
where $k_i$ is the degree of node $i$. This defines an unbiased random walk (URW) over the graph, assigning equal probability for transitions to happen from one node $i$ to its neighbours $j\in \p i$, which is known to be ergodic. 

\begin{figure}[t]
\includegraphics{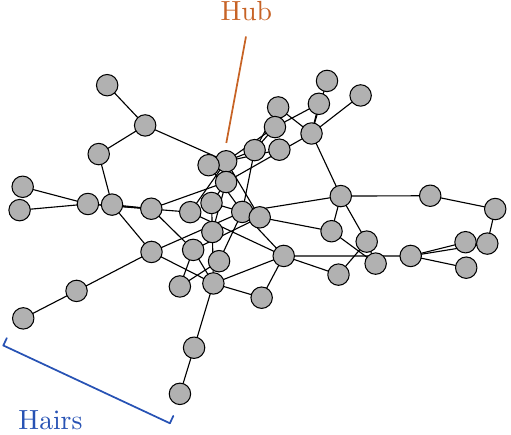}
\caption{ER graph of size $N=46$ obtained with $\kappa=3$ showing a hub node and two dangling chains or hairs.}
\label{fig1}
\end{figure}

Many dynamical observables can defined for the URW. The one that we consider is the mean degree
\be
C_n = \frac{1}{n}\sum_{\ell=1}^n k_{X_\ell},
\ee
visited by a trajectory of the URW over a time $n$. The large deviations of this observable have been studied recently \cite{bacco2015b,bacco2015,coghi2018b} and are interesting in that they show very different mechanisms creating large fluctuations of $C_n$ above the expected value $c^*=\kappa+1$ and small fluctuations of $C_n$ below $c^*$. Large degree fluctuations, on the one hand, come from rare trajectories of the URW that localise around a hub of highly connected nodes (see Fig.~\ref{fig1}a), resulting in an effective Markov chain for $s>0$ that concentrates on this hub as $s\ra\infty$. On the other hand, small degree fluctuations come from rare trajectories that localise on low-degree nodes and, for the smallest degree fluctuations, on ``dangling chains'' or ``hairs'' consisting of two nodes of degrees 2 and 1, respectively, resulting in $c_{\min}=1.5$ (see Fig.~\ref{fig1} and Fig.~2 in \cite{coghi2018b}).

These two fluctuations regimes are very different because large ER graphs typically have a single hub but many dangling chains. Moreover, dangling chains are not connected to each other, but are connected instead via high-degree nodes around the hub, so the small degree fluctuation regime, related to $s<0$, must come with a symmetry breaking mechanism, akin to a phase transition, whereby the URW localises on a particular dangling chain (out of many equivalent chains), which depends on the initial value of the random walk and the graph itself \cite{coghi2018b}.

\begin{figure*}[t]
\includegraphics{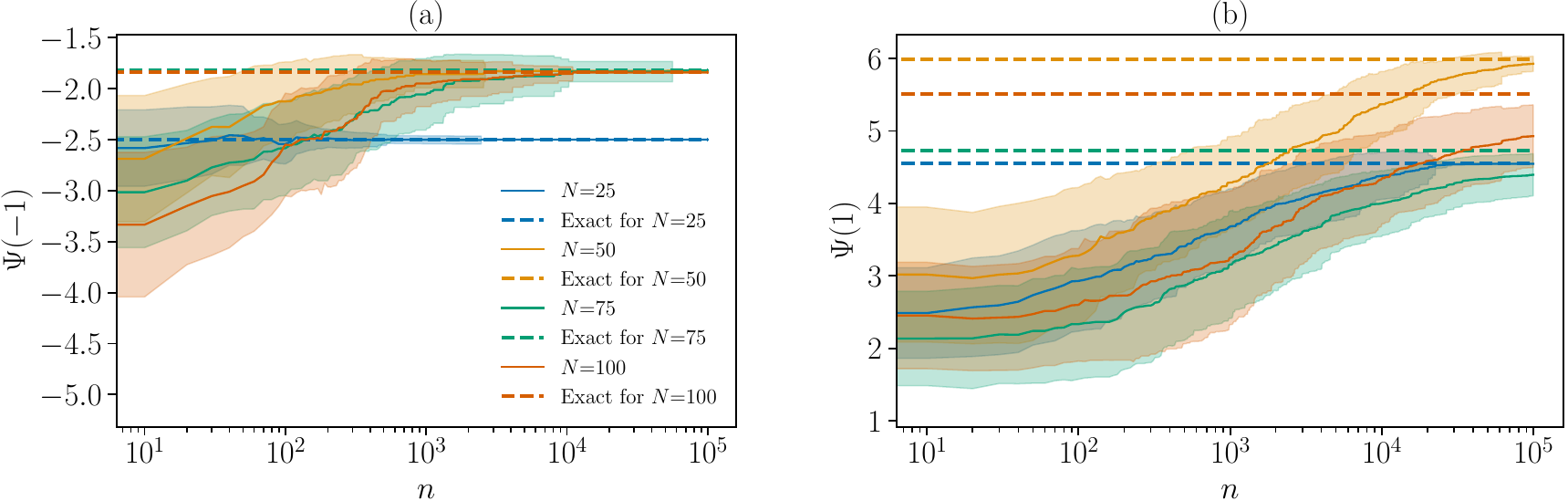}
\caption{Convergence of APM starting from random initial states for (a) $s=-1$ and (b) $s=1$. The colored lines show the average of the eigenvalue estimate $\hat\Psi_n(s)$ calculated over 100 simulations for different graph sizes (see legend) while the colored regions show the standard deviation. Learning rate: $\alpha=0.1$. The graph used is the one shown in Fig.~\ref{fig1}.}
\label{figscgf}
\end{figure*}

\begin{figure*}[t]
\includegraphics{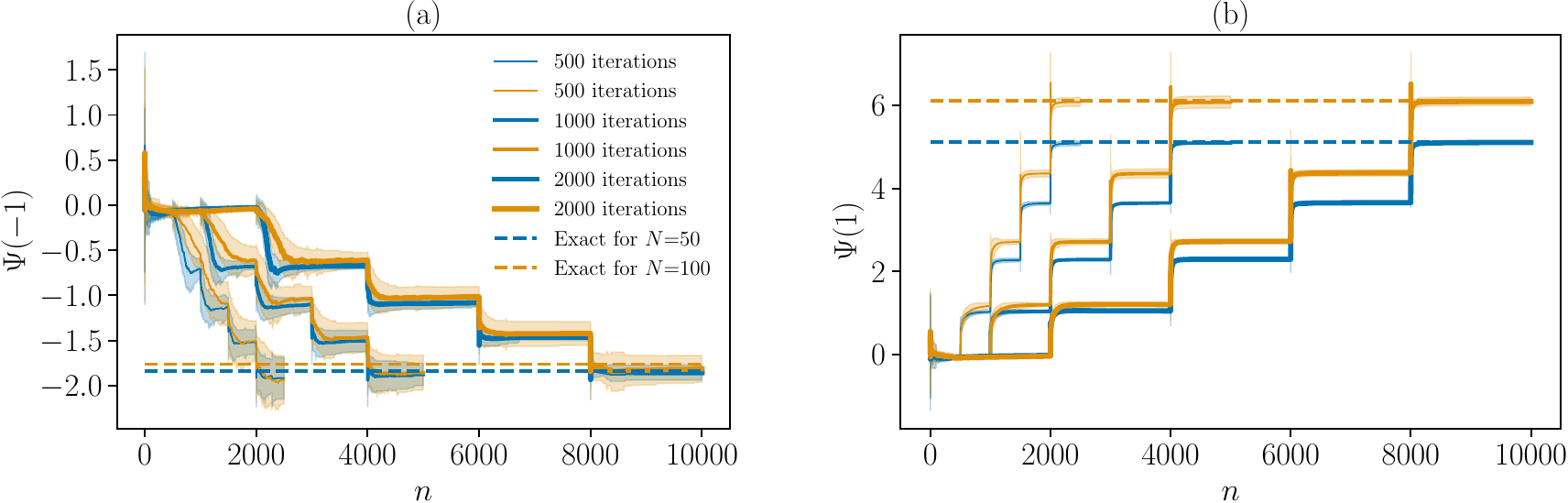}
\caption{Convergence of APM with transfer learning for (a) $s=-1$ and (b) $s=1$. The colored lines show the average of  $\hat\Psi_n(s)$ calculated over 100 simulations for two different graph sizes (see legend) while the colored regions show the standard deviation. Four steps were used to reach the final SCGF by changing $s$ in steps of $\Delta s=0.25$. Learning rate: $\alpha=0.1$.}
\label{figscgftransf}
\end{figure*}

This transition is especially visible when the connectivity parameter $\kappa$ is small and translates mathematically into a sharp corner of $\Psi(s)$ near $s=0$ and a near-linear branch of $I(c)$ left of the expected value $c^*$, corresponding to the zero of $I(c)$ (see Fig.~\ref{figldfct1} and Fig.~1 in \cite{coghi2018b}). These features of $\Psi(s)$ and $I(c)$ get more pronounced as the graph size $N$ increases, suggesting the existence of a DPT in the fluctuations of $C_n$, which should be first-order given that $\Psi'(s)$ develops a jump singularity. Whether this DPT is a ``true'' phase transition in the thermodynamic limit where $N\ra\infty$ is still unresolved. The numerical evidence published so far suggest that it is, but there is no rigorous proof yet that $\Psi(s)$ has a singularity in the thermodynamic limit \footnote{For a proof of the existence of DPTs in other types of graphs, see the recent work of Carugno, Vivo, and Coghi \cite{carugno2022b}.}.

The exact nature of the transition is not crucial for us; what is interesting is that a sharp transition or crossover is visible already for $N=50$, so one does not need to consider extremely large system sizes to study its effect on numerical computations of large deviation functions. For this reason, the URW on ER graphs is a good benchmark for testing the efficiency of numerical methods near DPTs and for comparing such methods. It is also a simple benchmark, as it does not involve many interacting particles. For an application of the APM to larger graphs, showing the efficiency of the algorithm on larger systems, we refer to the recent study of Di Bona \emph{et al.}~\cite{dibona2022}.

\subsection{Convergence of APM}

We show in Fig.~\ref{figscgf} the evolution of the SCGF returned by APM over time (i.e., iterations) for two values of $s$, namely, $s=-1$ and $s=1$, and different graph sizes. As can be seen, the convergence of $\hat\Psi_n(s)$ towards the exact value $\Psi(s)$, calculated by exact diagonalization \footnote{The large deviation functions have not been obtained analytically for this system \cite{coghi2018b}, so we rely on exact diagonalization results for comparison.}, is slow because the APM needs to learn by single-step exploration where the corresponding rare event takes place in the graph, leading to a long ``exploration'' phase starting from the initial position $x_1$, which is chosen to be a random node on the graph. The case $s=1$, for example, corresponds to a larger-than-average degree fluctuation, associated with paths of the random walk concentrating on the high-degree nodes, which need to be found and explored sufficiently in time before the eigenvalue can converge. This takes longer the bigger the graph is, as the random walk needs to reach the hub of the graph starting from random nodes, which are further from the hub, on average, the bigger the graph is. For $s=-1$, on the other hand, the random walk needs to reach hairs of the graph, a faster process, as shown in Fig.~\ref{figscgf}(a), since the number of hairs grows with the system size.

\begin{figure*}[t]
\includegraphics[width=\textwidth]{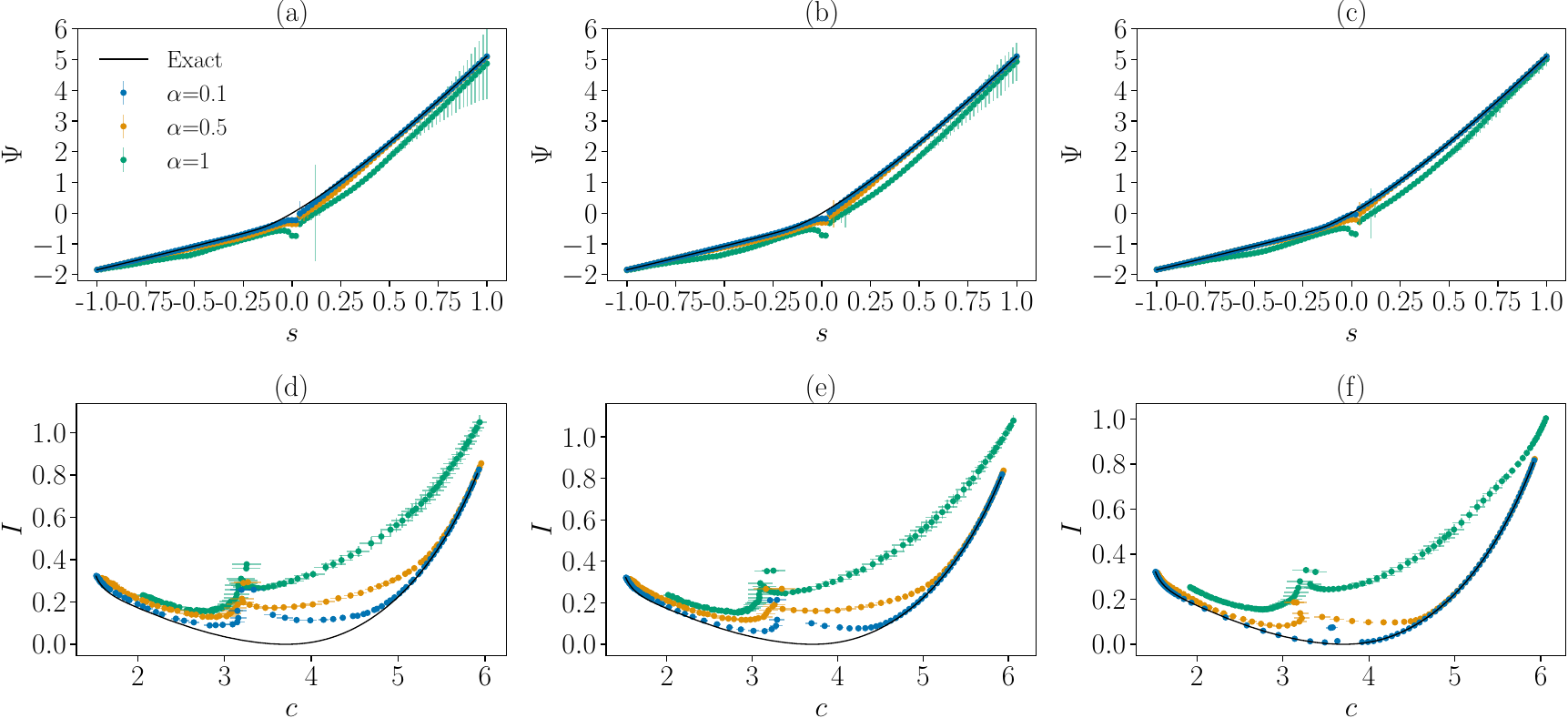}
\caption{Estimation of the SCGF and rate function with transfer learning for the graph shown in Fig.~\ref{fig1}. Top row: Estimated SCGF from the dominant eigenvalue for (a) 500, (b) 1000, and (c) 10 000 iterations per $s$ value, i.e., per transfer learning step. Botton row: Estimated rate function for (d) 500, (e) 1000, and (f) 10 000 steps per $s$ value.}
\label{figldfct1}
\end{figure*}

\begin{figure}[t]
\includegraphics[width=3.3in]{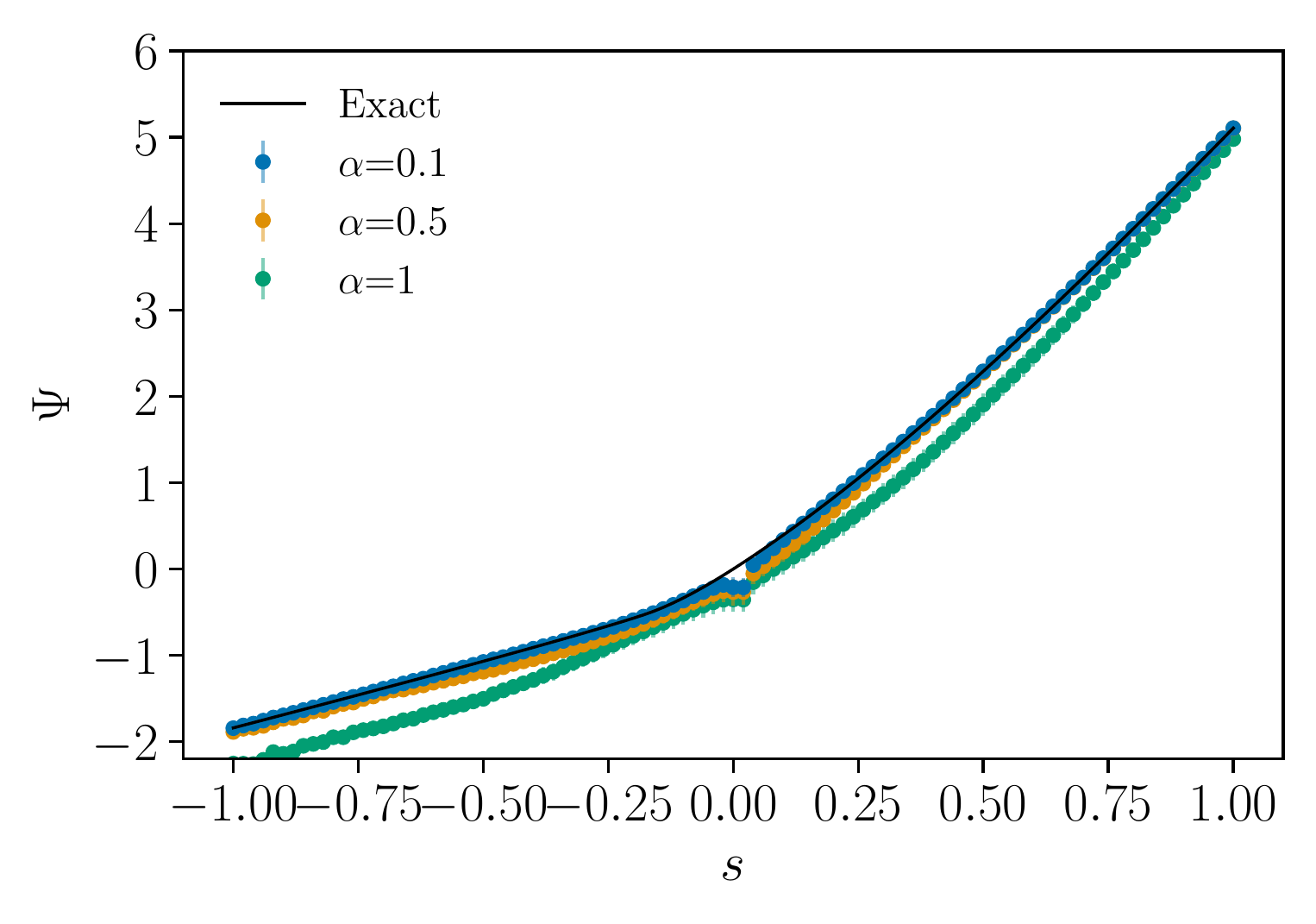}
\caption{Estimated SCGF with transfer learning using the additive estimator $\bar\Psi_n(s)$, defined in Eq.~\eqref{eq:psicostave1}, instead of the eigenvalue estimator $\hat\Psi_n(s)$. Parameters: 1000 iterations per transfer plateau spaced with $\Delta s=0.02$.}
\label{figscgfcost1}
\end{figure}

In practice, we should take advantage of the exploration phase, since it visits different regions of the graphs, associated with different degree fluctuations and, therefore, different $s$ values. Thus, instead of trying to reach the correct eigenvalue for a given $s$, we should start the algorithm at a small value of $s$ close to $s=0$, let it run for this value, which is only a small perturbation of the result $\zeta_0 =1$, and repeat the process by incrementing or ``quenching'' $s$ in small steps of $\Delta s$ \cite{ferre2018,yan2022}, re-initializing the annealing sequence at each step. The result with $\Delta s=0.25$ is shown in Fig.~\ref{figscgftransf} again for $s=-1$ and $s=1$, two graph sizes, and different number of iterations for each quench level. As can be seen, the APM now converges quickly for each value of $s$ considered, leading to a much smaller error region (shaded region) compared to the results of Fig.~\ref{figscgf}, especially for $s=1$. Moreover, the convergence now does not depend significantly on the system's size, especially again for $s=1$, which shows quenched plateaux being reached more or less at the same times for the two sizes considered ($N=50$ and $100$). This arises because the simulation for a given $s$ ``locally'' informs the next simulation for $s+\Delta s$ by transferring the state of the Markov chain and the computed eigenvector from one simulation to another.

\begin{figure*}[t]
\includegraphics[width=0.9\textwidth]{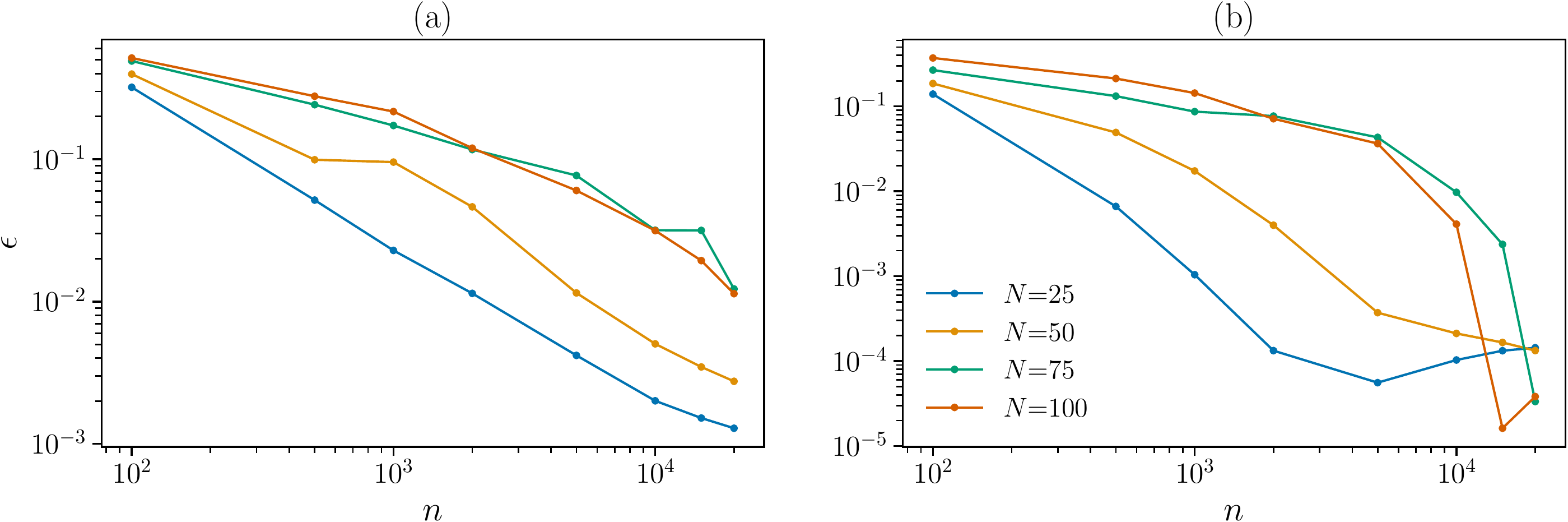}
\caption{Relative error of $\bar\Psi_n(s)$ for (a) $s=-1$ and (b) $s=1$ as a function of the number of iterations per step of the transfer learning. Parameters: $\alpha=0.1$ and $\kappa=3$.}
\label{figerr1}
\end{figure*}

We call this transfer of information between different runs of the algorithm ``transfer learning'' in analogy with a similar method used in machine learning for transferring parameters from one learned model to another. An obvious advantage of this method is that we are able to efficiently obtain the SCGF for a range of $s$ values by using a small step $\Delta s$, as shown in Fig.~\ref{figldfct1}(a)-(c), which was obtained for the graph shown in Fig.~\ref{fig1} using $\Delta s=0.02$. In this figure, we also show the effect of using different iterations for each run, steps or ``quenches'' of $s$, as well as different learning rates $\alpha$. The results show overall that convergence with small error bars is achieved with 1000 iterations, and lower when considering the tails of the SCGF. What is more important for convergence is the learning rate, which needs to be low enough so that the learning is annealed in a slow way, favouring exploration. By trying different values, we have found that $\alpha=0.1$ gives good results for the graph sizes considered. Above this value, the APM is annealed too quickly, leading to a negative bias in the SGCF, especially near the DPT around $s=0$, as shown in Fig.~\ref{figldfct1}.

This bias is seen more clearly at the level of the rate function, shown in the lower plots of Fig.~\ref{figldfct1}. There we see that the APM gets trapped if annealed too quickly in different phases of the DPT, corresponding to different regions of the graphs that either have a low or a high connectivity, resulting in a positive bias in the estimated rate function. 

The bias is reminiscent of hysteresis effects seen in Monte Carlo simulations of equilibrium systems having first-order phase transitions, and can be alleviated by performing longer simulations and, in our case, by using a low learning rate. Convergence around the DPT can also be improved by using a dynamical variant of the replica exchange method \cite{swendsen1986}, whereby two simulations with different $s$ values are randomly exchanged, as proposed and tested recently \cite{yan2022}. We have not used this method for APM, but we think that it might be useful when dealing with very large systems showing DPTs.

It is important to note again that $I(c)$ is estimated from the additive cost $K_{n,s}$ shown in \eqref{eq:costest1} and not from the Legendre transform of $\hat\Psi_n(s)$, since the latter is not necessarily convex, as seen from Fig.~\ref{figldfct1}. The SCGF itself can also be estimated as a time-average, as noted before, using the estimator $\bar\Psi_n(s)$ defined in \eqref{eq:psicostave1}.  This method in fact converges faster and is more stable, as shown in Fig.~\ref{figscgfcost1}, and is therefore our preferred method for computing the SCGF. Using the dominant eigenvalue gives good results, but the additive estimator gives better results for the same number of iterations and learning rate because it is a self-averaging quantity. 

To obtain an idea of the error associated with the additive estimator, and therefore of APM in general, we have computed $\bar\Psi_n(s)$ for $s=-1$ using four levels or quenches separated by $\Delta s=0.25$ and have computed the relative error $\epsilon$ between the final value of the estimator and the exact SCGF obtained by direct diagonalization. The results, also computed for $s=1$, are shown in Fig.~\ref{figerr1} as a function of the number $n$ of iterations per quench and for different system sizes. As expected, the relative error decreases with $n$ in a way that appears roughly to be polynomial with $n$ for all system sizes, demonstrating the efficiency of APM coming from the importance sampling step, which has the effect of biasing the random walk in the relevant fluctuation region regardless of system size. From the plot, we can also see that the error is smaller for $s=1$, confirming the smaller error bars shown in Fig.~\ref{figscgftransf} for this value of $s$. 

An error analysis can also be performed, in principle, for the eigenvalue estimator $\hat\Psi_n(s)$ of the SCGF, although in this case one must decide whether to compare the final value estimated or average over an entire simulation. We do not show results for this estimator, as they are less stable compared with the additive estimator $\bar\Psi_n(s)$. 

The error bars shown in all the plots discussed so far were obtained by repeating runs of the APM algorithm, but it is important to note that error bars could have been obtained for the additive estimators from single runs using batch mean methods \cite{asmussen2007}. This, in fact, is a further advantage of using additive estimators for the SCGF and the rate function. Defining error bars for the eigenvalue estimator based on single runs is more difficult, since this estimator is not a time average, but an annealed version of the eigenvalue returned by APM. As far as we know, no theoretical work has been done on deriving a central limit theorem for the eigenvalue estimator, which would serve as the basis for defining error bars for this estimator. Our advice again, based on the results that we have obtained, is to use additive estimators whenever possible. They converge quicker and have better variance properties compared to the eigenvalue estimator.

\subsection{Effective Markov chain}

\begin{figure*}[t]
\includegraphics[width=0.9\textwidth]{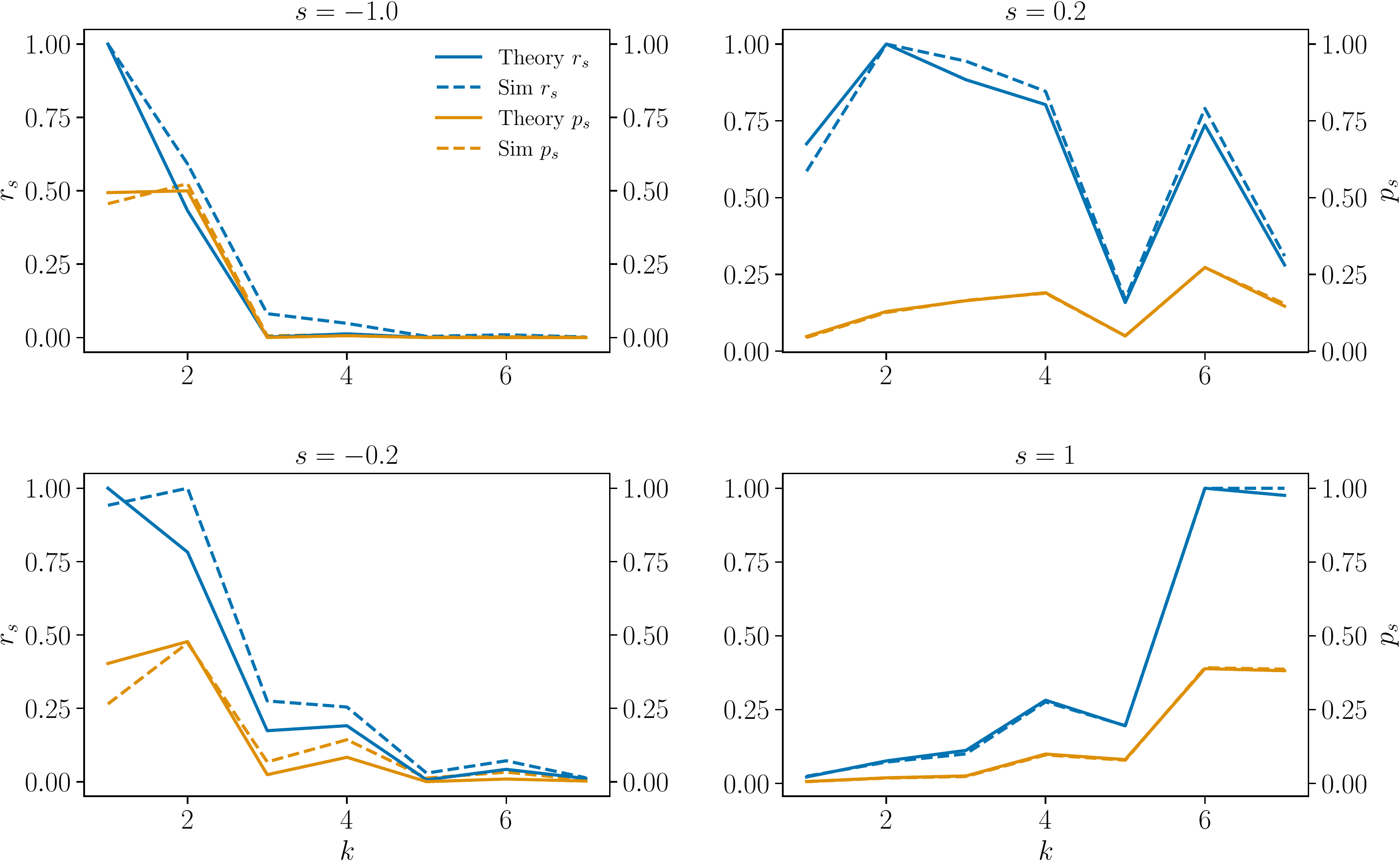}
\caption{Dominant eigenvector $r_s$ and stationary distribution $p_s$ of the effective Markov chain plotted as a function of node degree. The numerical result, averaged over 100 simulations, are compared for different $s$ with theoretical results obtained by direct diagonalization. Learning rate: $\alpha=0.1$. The graph used is the one shown in Fig.~\ref{fig1}.}
\label{figps1}
\end{figure*}

The APM yields not only the SCGF and rate function, but also the effective process associated with the set of trajectories creating a given fluctuation $C_n=c$ in the long-time limit. This process was discussed extensively in an earlier study \cite{coghi2018b}. In Fig.~\ref{figps1}, we show the stationary distribution $p_s$ of this process obtained by running the APM for some time after it has found the SCGF. The results are plotted, together with the dominating eigenvector $r_s$, for four $s$ values located around the DPT and compared with the theoretical results obtained again by direct diagonalization. Note that, for convenience, we plot $r_s$ and $p_s$ as a function of the node degree $k$ rather than as a function of the node state, although both are of course functions of $i\in\Gamma$ \footnote{The mapping from component to degree is done simply by adding the components of the eigenvector with the same degree.}. Moreover, we normalized $r_s$ in such a way that $\max_k r_s(k) =1$.

The plots show that APM is able to recover $r_s$ and $p_s$ with good accuracy. We recall that $r_s$ is used in the importance sampling step to bias the random walk in the region where a given degree fluctuation is likely to take place, while $p_s$ shows where that region is. In the case $s=-1$, for example, which corresponds to small degree fluctuations, the plot of $p_s$ shows that the effective random walk concentrates on small degree nodes (\emph{viz.}, on hairs), whereas for $s=1$ the random walk concentrates on higher degree nodes located around the hub of the graph. We refer again to \cite{coghi2018b} for a detailed discussion of this concentration and its relation to the DPT. 

What is important to note in relation to the APM is that, for the purpose of biasing the process in the importance sampling step, one does not need to estimate the dominant eigenvector $r_s(i)$ with high precision for all states $i$ of the state space $\Gamma$, but only for those states that are most often visited, as determined by $p_s$. This arises because the additive estimators of the SCGF and the rate function are in fact estimators of expectations with respect to $p_s$ \cite{chetrite2015}. As a result, we do not need in practice to represent $r_s$ over the whole of $\Gamma$, but can truncate that representation to states that have been visited a certain number of times, in analogy with real-space renormalization methods \cite{gorissen2009}. This can be done dynamically within simulations to save memory (as done, e.g., in reinforcement learning algorithms \cite{sutton2018}), opening the possibility of applying APM to very large systems.

\section{Concluding remarks}

We have tested in this paper the APM algorithm on a simple random walk to show the efficacy of this algorithm for computing large deviation functions characterizing the fluctuations of time-additive functions of Markov processes. Compared to previous studies on this algorithm \cite{borkar2004,ahamed2006,basu2008,ferre2018}, we have studied the effect of the learning rate on the convergence of the algorithm and demonstrated that the algorithm is not slowed down around DPTs when a relatively small learning rate is used. In practice, different learning rates (starting from large to small), initial conditions, and simulation times should be used to test whether APM converges in a robust way. One advantage of estimating the rate function using the cost estimator is that the value obtained is an upper bound on the true rate function (within statistical errors), so reducing that cost with different parameters necessarily leads to an improvement in the estimated function, as illustrated in Fig.~\ref{figldfct1}.

Following \cite{ferre2018}, we have also shown that the SCGF and rate function are most efficiently computed by performing long simulations in which the large deviation parameter $s$ is increased in small steps, using the knowledge of the computed eigenvalue and eigenvector at each step to guide with importance sampling the random walk in the fluctuation region of interest. The use of importance sampling is critical for the algorithm to work -- indeed for any large deviation simulation method to work -- and leads APM to gradually learn the effective process, which is known to be efficient for sampling large deviations and rare events in general \cite{asmussen2007,bucklew2004,guyader2020}. 

The effective process can be estimated or learned in other ways, for instance, using splitting or cloning algorithms \cite{nemoto2016}, stochastic optimization \cite{yan2022}, or reinforcement learning \cite{oakes2020,rose2021,das2021}. Compared to these approaches, APM has some advantages:
\begin{enumerate}
\item The algorithm is guaranteed to converge to the correct SCGF \cite{borkar2004}, and so has no bias, essentially because the feedback mechanism that updates the importance sampling is based on the power method. Other methods based on variational representations of the SCGF (see, e.g., \cite{yan2022}) are generally found to converge, but it is often difficult to guarantee that the solution they return is the optimal solution (global minimum) corresponding to the SCGF. 

\item The annealing of the stochastic approximation of the eigenvector has the effect of stabilizing the adaptive importance sampling, as it gradually reinforces previous updates and filters out the randomness or noise inherent in the new generated states. Importance sampling can also be included and adapted in cloning but the feedback mechanism that has been proposed so far \cite{nemoto2016} is not annealed and has not been proved to converge.

\item Since the estimators of the SCGF and rate function are time-additive quantities, just like the observable itself, their statistical errors can be computed directly in single runs of the algorithm using standard batch mean methods \cite{asmussen2007}. The same applies to stochastic optimisation and reinforcement learning approaches, but not to cloning, which is not based on additive estimators. 
\end{enumerate}

More work is needed to compare these methods and the many more that have been proposed recently for computing large deviation functions and, moreover, to develop good benchmarks for comparing different methods. In a sense, all methods should have the same complexity, if implemented efficiently, since they are aimed at solving the same problem -- that of finding a dominant eigenvalue and its eigenvector \cite{chetrite2015}. However, some methods might be easier to implement, depending on the process and observable considered. Moreover, some methods might be advantageous in terms of stability, for proving convergence, or for computing errors, as discussed here.

\begin{acknowledgements}
F.C.\ is grateful to Giorgio Carugno for interesting discussions. Part of the computations were carried on the Apocrita HPC facility at Queen Mary University of London, supported by QMUL Research-IT.
\end{acknowledgements}

\bibliography{masterbibmin}

\end{document}